\documentclass[aip,rsi,preprint]{revtex4-1}

\usepackage{amsmath,graphicx,epsfig}

\begin{document}

\title{Sympathetic and swap cooling of trapped ions by cold atoms in a MOT}
\author{K. Ravi,$^1$ Seunghyun Lee,$^1$ Arijit Sharma,$^1$ G. Werth$^2$ and S. A. Rangwala}
\email {sarangwala@rri.res.in}
\affiliation
{$^1${Raman Research Institute, Sadashivanagar, Bangalore 560080, India}\\
$^2${Institut f\"ur Physik, Johannes-Gutenberg-Universit\"at, D-55099 Mainz, Germany}}
\setcounter{secnumdepth}{-1}

\date{\today}

\begin{abstract}

{\bf A mixed system of cooled and trapped, ions and atoms~\cite{Smi05,Cet07,Gri09,Zip10a,Sch10,Rav11}, paves the way for ion assisted cold chemistry~\cite{Wil08,Zip10b,Hal11,Rel11} and novel many body studies~\cite{Cot02}. Due to the different individual trapping mechanisms, trapped atoms are significantly colder than trapped ions, therefore in the combined system, the strong binary ion$-$atom interaction results in heat flow from ions to atoms. Conversely, trapped ions can also get collisionally heated by the cold atoms, making the resulting equilibrium between ions and atoms intriguing. Here we experimentally demonstrate, Rubidium ions (Rb$^+$) cool in contact with magneto-optically trapped (MOT) Rb atoms, contrary to the general expectation of ion heating for equal ion and atom masses~\cite{Maj68}. The cooling mechanism is explained theoretically and substantiated with numerical simulations. The importance of resonant charge exchange (RCx) collisions, which allows swap cooling of ions with atoms, wherein a single glancing collision event brings a fast ion to rest, is discussed.}

\end{abstract}

\maketitle

This letter investigates energy transfer from trapped $^{85}$Rb$^+$ ions to laser cooled $^{85}$Rb atoms in a MOT, which have equal masses, resulting in ion cooling. Since Rb$^+$ ions have a closed shell electron configuration they are not amenable to direct laser cooling. In addition, trapped ions heat~\cite{Maj05} due to factors such as trap imperfections, background gas collisions and radiofrequency (RF) heating due to ion$-$ion repulsion. In the present experiment, collisions with cold atoms is the only available cooling channel for Rb$^+$ ions. To explain the collisional ion cooling we first discuss the experimental arrangement, then furnish the theoretical argument for efficient cooling of ions by MOT atoms, followed by computational results which expand the scope of the binary ion$-$atom collision. The experimental observation of ion cooling and trapped ion number equilibrium by the MOT atoms is then presented and discussed. 

\begin{figure}
\includegraphics[width=8.5 cm]{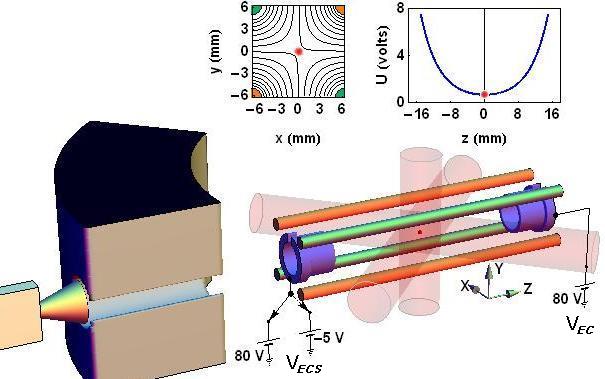}
\caption{{\bf Experimental Schematic.} The nested MOT in the linear Paul trap with CEM is shown. The ion$-$atom combined trap is constructed, so as to coincide the MOT center with the minimum of the ion trap secular potential, at the origin. The MOT is formed at the intersection of the six cooling and repump beams. Ions are created from the excited MOT atoms, by absorption of a photon fom a blue light source (not shown). The out of phase oscillating voltage on the quadrupole electrodes (see x$-$y contour plot) effects x$-$y trapping, while a constant end electrode voltage V$_{EC}=80$ V confines along the z direction (see z potential). The size of the MOT is illustrated in red in the cut views of the ion trap potential, demonstrating V$_{MOT}\ll$V$_{IT}$. The confined ions are detected by a CEM (housed in an extension to the main chamber), post extraction from the trap, by switching V$_{ECS}$ between $80/-5$ V as shown. The port hole in the outer wall of the chamber constitutes a drift region for time-of-flight measurement of the ions extracted.}
\label{Fig:ExperimentSchematic}
\end{figure}

The experimental schematic is illustrated in Fig.~\ref{Fig:ExperimentSchematic} and the experimental technique~\cite{Rav11} and operational details are briefly described in the Methods section. As seen in the figure, the linear ion trap volume (V$_{IT}$) is much larger than the volume of the cold atoms in the MOT (V$_{MOT}$). The saturated MOT with $2.3(\pm0.2)\times 10^6 $ atoms has a full width at half maximum (FWHM) of $\approx 1$ mm and is formed at the center of the ion trap secular potential (origin). Ions are created by two-photon ionization from the MOT (see Methods) with negligible recoil.  The ions are detected by a channel electron multiplier (CEM) and amplifier assembly, which converts the incident extracted ions from the trap to a proportional voltage signal.

To demonstrate the cooling of trapped ions by collision with MOT atoms, we closely follow the early, seminal work of Major and Dehmelt~\cite{Maj68}. 
In describing the ion$-$atom collisions, the MOT atom temperature of $\approx$100 $\mu$K permits the setting of atom velocity, $\mathbf{v_A}=0$.
For initial ion velocity $\mathbf{c}$, the post collision velocity of the ion $\mathbf{c'}$ is given by 
\begin{equation}
\mathbf{c'} = (m_{A}/M) c ~\hat{\mathbf{\theta_{c}}}+ (m_{I}/M) \mathbf{c},
\label{Eqn:first}
\end{equation}
where $\hat{\mathbf{\theta_{c}}}$ is a unit vector at an angle $\theta_{c}$ with respect to $\mathbf{c}$, $m_I$ is the ion mass, $m_A$ the atom mass and $M=m_I+m_A$. The ion motion can be decomposed into its macromotion and its in-phase micromotion oscillation with the applied electric field~\cite{Zip11}. The pre and post collision ion velocities then are, $\mathbf{c} = \mathbf{u}+\mathbf{v}$ and $\mathbf{c'} = \mathbf{u'}+\mathbf{v}$, where $\mathbf{u}$ and $\mathbf{u'}$ are the respective secular velocities, $\mathbf{v}\propto\mathbf E sin(\phi_{RF})$, is the micromotion velocity and $\phi_{RF}$ is the phase of the electric field $\mathbf E$ at the instant of collision. It is the reduction in the average secular motion velocity, $ \left\langle\left|\mathbf u\right|\right\rangle$ with collisions (time), that leads to ion cooling (see the Methods section). All collisions with the MOT atoms occur close to the origin, where $\left\langle \left|\mathbf v\right|\right\rangle\rightarrow 0$, leading to ion cooling irrespective of $m_I/m_A$ due to a reduction of $\left\langle \left|\mathbf u\right|\right\rangle$. The average cooling efficiency per collision is maximum for $m_I=m_A$. It is therefore the spatially compact density of the MOT which leads to the collisional cooling of the ions.

The analysis above, is valid for elastic scattering. For Rb$^+-$Rb collisions the RCx channel~\cite{McD64,Smi03,Cot00,Smi01} plays a key role in cooling the Rb$^+$ ions. In RCx, the atomic valence electron transfers from atom to ion with no change in the dynamical or internal states of the colliding partners, apart from swapping their respective charge states. For the ion$-$atom collision energy ($E$) involved in the experiment, the RCx cross section $\sigma_{cx}\propto 1/E^{1/2}$, is comparable to the ion-atom elastic cross section~\cite{Cot00,LLQM} $\sigma_{el}\propto 1/E^{1/3}$, so both channels participate in ion cooling. In those glancing collisions where RCx occurs, the swapping of the charge state results in an ion at rest. This swap cooling occurs preferentially at the ion trap minimum, where the MOT density is maximum. Since glancing collisions are overwhelmingly more probable than head-on collisions, the RCx mechanism for transferring energy from ions to atoms dominates the elastic channel. The difference in the  evolution of ion cooling by elastic and RCx processes, in a multiple scattering framework, is brought out in the numerical simulations discussed below.

\begin{figure}
\includegraphics[width=8.5 cm]{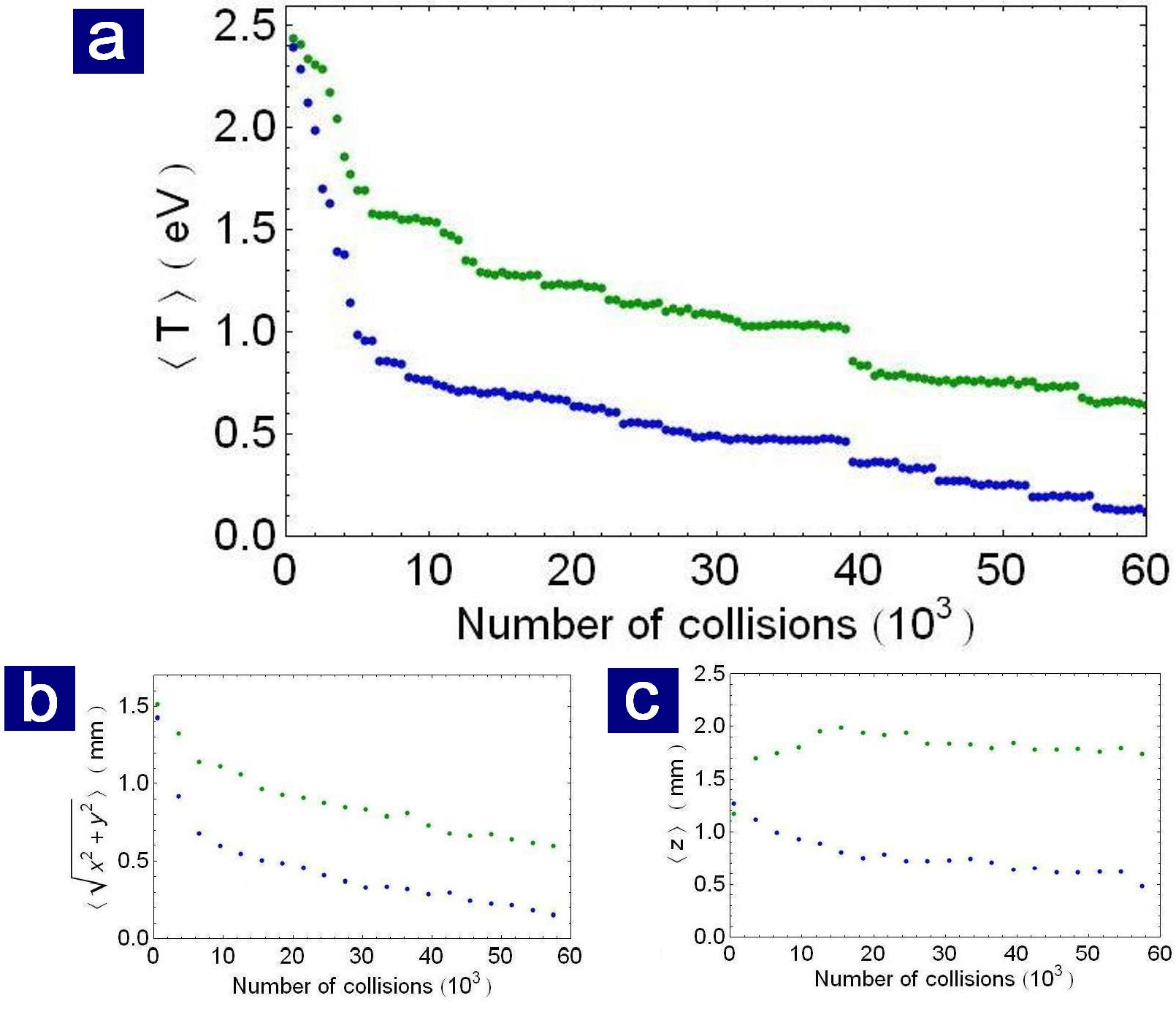}
\caption{{\bf Ion Cooling Simulation Results.} Here green represents elastic and blue elastic$+$RCx collisions. Panel (a) illustrates the fall in average ion kinetic energy ($\left\langle T \right\rangle$) as a function of the total number of collision events (which corresponds directly to evolution time) for the 100 independent ions with the localized atoms. In both cases $\left\langle T \right\rangle$ decreases with collision number, however for the elastic$+$RCx collision, the reduction in $\left\langle T \right\rangle$ is much faster when the ions are more energetic. Reduction of the average displacement for the ions, with collision number in the $x-y$ and $z$ directions, are shown in panels (b) and (c), directly establishing the link between ion cooling and the reduction in the spatial spread of the ions. In the elastic case the initial reduction in transverse spread increases the longitudinal spread, which then slowly starts to reduce.}
\label{Fig:KinEng}
\end{figure}

Trajectories of 100 non-interacting trapped ions are computed, each undergoing multiple collisions within a localized spherical density distribution ($\rho_A=\rho_0$, a constant for $r\leq0.2$ mm and $\rho_A=0$ for $r>0.2$ mm) of atoms about the ion trap center. Elastic~\cite{McD64,CRCHBk} and RCx~\cite{Hol52,Ols69} collisions are both incorporated in the simulation (see the Methods section). The mean ion kinetic energy is shown in Fig.~\ref{Fig:KinEng}(a), and mean position displacement of the ion ensemble in Fig.~\ref{Fig:KinEng}(b) and (c). Ion cooling is clear from the reduction of mean kinetic energy per ion and the narrowing of the spatial spread of the non-interacting ions in the trap with time. The step changes in the kinetic energy occur for either a head on elastic or glancing RCx collision at the trap bottom. The ions cool much faster when the RCx channel is active along with the elastic one. The atom density is kept constant at the trap center to emphasize the role of compactness in the distribution of atoms for ion cooling. Naturally a compact gradient distribution of the atoms further enhances ion cooling.  

\begin{figure}
\includegraphics[width=8.5 cm]{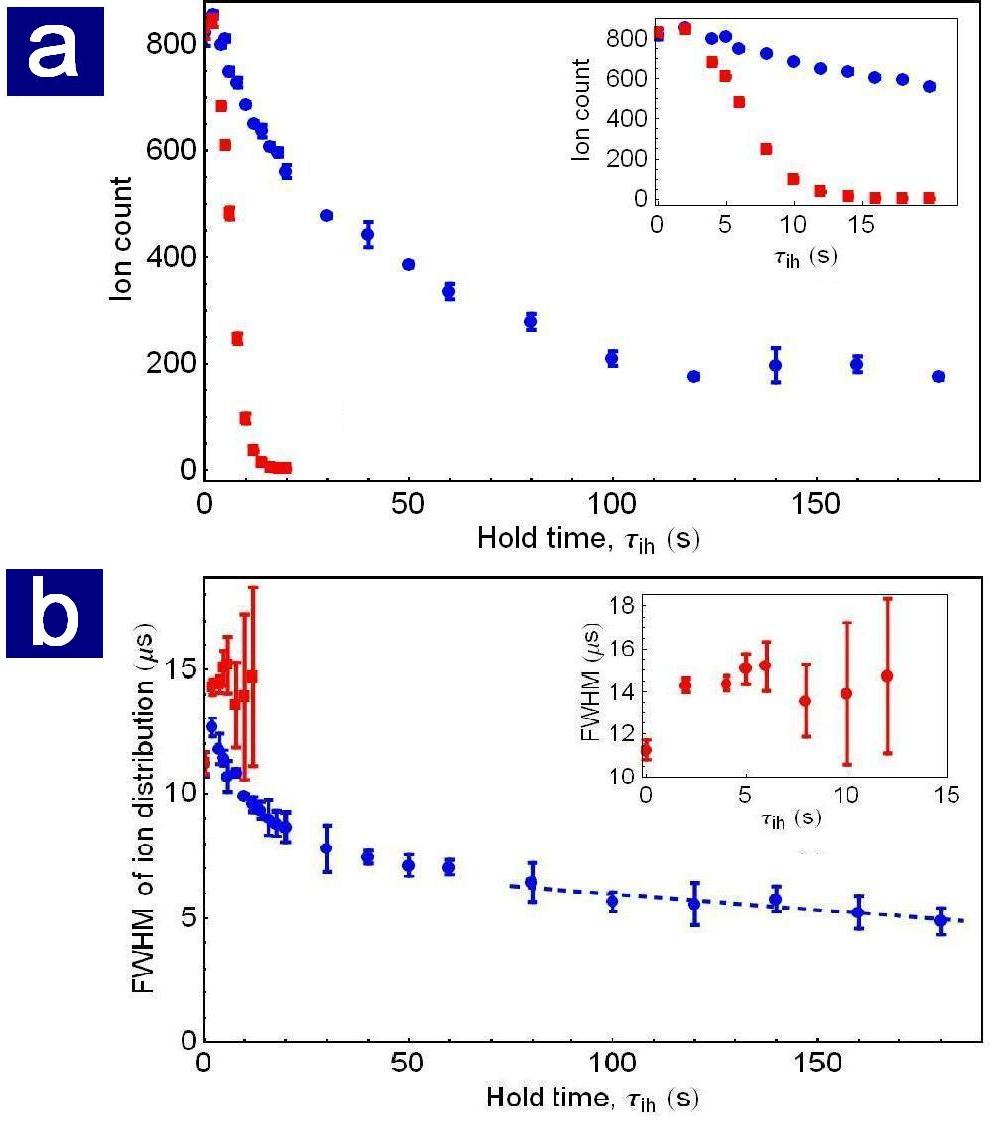}
\caption{{\bf Experimental Demonstration of Ion Cooling.} Case (1) no MOT is represented in red and case (2) with MOT in blue. Panel (a) plots the number of Rb$^+$ ion counts as a function of $\tau_{ih}$. Without a MOT, the ions exit the trap rapidly, while with cold atoms the ion loss is much slower and a stable number of ions ($ 187 \pm 9$) is trapped without detectable loss beyond $\tau_{ih}\geq 2$ minutes. Panel (b) illustrates the variation of the FWHM  of the ion ToF distribution against $\tau_{ih}$ for the two cases. For case (1) the FWHM increases in time as the trap empties out, while for case (2) a systematic decrease in the ion ToF distribution is seen, consistent with ion cooling. For $\tau_{ih}\geq 2$ min, when the trapped ion number has stabilized, the ToF width is still decreasing indicating continued ion cooling, as illustrated by a least square fit to the last six data points. The insets serve to illustrate case (1) data with clarity. The statistical standard deviation error bars are shown.} 
\label{Fig:Lifetime}
\end{figure}

Experimentally, ion cooling by cold atoms is implemented as follows. The Rb dispenser and the gradient magnetic field are ON throughout the measurement. The cold atoms are loaded (emptied) from the MOT, by switching the atom cooling and repump lasers ON (OFF), with a mechanical shutter. Initially, while the MOT is loaded for $\tau_{ml}=40$ s to saturation, the ion trap RF field is OFF. For ion loading, the ionizing blue light is briefly pulsed ON for $\tau_{il}=1$ s, simultaneously with switching ON the ion trap RF field for the remainder of the experimental cycle. The ions are trapped for a hold time, $\tau_{ih}$, following which they are extracted onto the CEM by switching V$_{ECS}$, as shown in Fig.~\ref{Fig:ExperimentSchematic}. The CEM ion count is measured as a function of $\tau_{ih}$ for two cases (1) without MOT atoms and (2) with MOT atoms. 

The CEM ion counts vs. $\tau_{ih}$ is plotted in Fig.~\ref{Fig:Lifetime}(a) and the FWHM of the ion arrival time-of-flight (ToF) distribution in Fig.~\ref{Fig:Lifetime}(b). For case (1) all the ions exit the trap by $\tau_{ih}\approx 15$ s and the ToF FWHM increases rapidly. 
In case (2) for $\tau_{ih}\geq 1$ s, the ion loss rate from the trap drops and so does the FWHM of the ion ToF distribution. 
Beyond $\tau_{ih}\geq 2$ minutes, the number of trapped ions stabilizes to a constant value, while the width of the ToF distribution nominally decreases, indicating that ion cooling is still underway. We therefore conclude that a number equilibrium between daughter ions and laser cooled parent atoms is achieved and a thermally stable equilibrium exists within experimental means. Finally since the cooling rate of the ion must overcome the heating rate for long term trapping, the lower bound on the initial cooling rate per ion is $dQ/dt|_{\tau_{ih} = 0}= - 0.038$ eV/s, as derived from fitting the heating observed in the MOT off measurement.

In our explanation of the above ion cooling we have exclusively focussed on the binary ion$-$atom interaction. Within this restriction we demonstrate that the ions are effectively cooled by collision with the localized, cold MOT atoms. It is the continuous atom cooling, which indirectly bleeds away energy from the trapped ions, and the constant number of atoms in the reservoir (MOT) that has the capacity to cool the ions. Without these two features simultaneously in place, the trapped ions will empty the atom trap very quickly, because of ion heating by various mechanisms and the relatively strong $-\alpha/r^4$ ion$-$atom interaction potential. The effect of ion$-$ion interactions on the cooling and the stabilization of the ion numbers has not been theoretically addressed here. The existence of ground and excited state populations within the cold atoms, with different polarizabilities, has also been ignored in the discussions. While these have an important part to play in the determination of the final state of the system, these details are not of primary relevance to the binary collisional cooling principle demonstrated here.

In conclusion we have shown that localized parent atom $-$ daughter ion collisions allow a viable ion cooling technique. This is an important development in understanding collisional cooling of ions and leads to intriguing questions for future study. For instance, what is the ultimate temperature to which the ions can be cooled by atoms?  What is the equilibrium state and configuration of the cooled ion$-$atom system? How are collective oscillations of the ions damped in this RCx active system? Most significantly, the stabilization of the cold ion$-$atom system sets the stage for ion involved production of cold molecules and generalization to studies with multiple species.

\section{Methods}

{\bf Experiment Construction and Operation}

The Rb MOT is vapour loaded by heating a Rb source. The MOT has six independent laser beams of 1 cm diameter and a gradient magnetic field of 12 Gauss/cm. Measurements on the cold atoms are made by flourescence detection and optical imaging. The Rb$^+$ ions are created from the trapped MOT atoms by two photon ionization, where one photon is available from the cooling laser itself. The second photon (456 nm), is emitted by a light emitting diode (LED), which is switched on to create ions from the excited MOT atoms. The ion trap is a linear Paul trap with hollow cylindrical end caps as shown in Fig.~\ref{Fig:ExperimentSchematic}. A sinusoidal voltage of amplitude $V_{RF}=91$ V, frequency $\Omega_{RF}=600$ kHz with $180^\circ$ phase difference between adjacent rods is applied for x-y confinement. Confinement in z is effected by a DC voltage $V_{EC}=80$ V applied to end electrodes along the z-axis. These ions are destructively detected by sweeping out the trapped ions through the hollow end electrode and a drift region, onto a CEM, by changing the voltage, $V_{ECS}$ from 80 V $\rightarrow-5$ V. 

{\bf Ion Cooling due to Atom Localization}

The change in ion temperature on undergoing collisions is principally due to a change in its $\left\langle \mathbf{u}\right\rangle$, as a collisional change in  position dependent $\mathbf{v}$ results only in the change of phase of the micromotion. 
Substituting $\mathbf{c} = \mathbf{u}+\mathbf{v}$ and $\mathbf{c'} = \mathbf{u'}+\mathbf{v}$ in Eqn.~\ref{Eqn:first} and with some regrouping of the terms, we obtain an expression for $u'^{2}-u^{2}$ as
\begin{equation}
u'^{2}-u^{2}=-2 m_{I}m_{A}(u^{2}+2 \mathbf{u} \cdot \mathbf{v}+v^{2})(1-\cos \theta_{c})/M^{2}
+2 m_{A}(v^{2}+\mathbf{u \cdot v}-c~\hat{\mathbf{\theta}}_{c}\cdot \mathbf{v})/M.
\label{Eqn:second}
\end{equation}
Conventionally, buffer gas floods the entire ion trap volume, allowing the approximation $c~\hat{\mathbf{\theta}}_{c}\cdot \mathbf{v}\approx v^{2} \cos \theta_{c}$ as most collisions occur close to the classical turning points of the macromotion where the ion spends most of its time and the micromotion velocity dominates, i.e. when $\left\langle \left|\mathbf u\right|\right\rangle\rightarrow 0$ and $\left\langle \left|\mathbf v\right|\right\rangle\gg 0$. This results in ion cooling for $m_A<m_I$, ion heating for $m_A>m_I$, and no net change of ion temperature for $m_A= m_I$. In practice effective ion cooling is seen when $m_A \ll m_I$, because of ion heating mechanisms. 

The present case however represents V$_{MOT}\ll$V$_{IT}$, where the cold atoms are localized at the ion trap center, i.e when $\left\langle \left|\mathbf u\right|\right\rangle\gg 0$ and $\left\langle \left|\mathbf v\right|\right\rangle\rightarrow 0$. Evaluating Eqn.~\ref{Eqn:second} above in this limit and after carefully taking a time average over $\phi_{RF}$, 
we obtain the expression for $\left\langle u'^{2}-u^{2}\right\rangle$ as,
\begin{equation}
\lim_{v\to0}\langle u'^{2}-u^{2}\rangle \approx - 2m_{I}m_{A}\langle u^{2}\rangle(1-\cos \theta_{c})/ M^{2},
\label{Eqn:third}
\end{equation}
which ensures a reduction in the average macromotion energy of the ion per collision and that maximum cooling in a collsion, for any particular deflection angle, occurs for $m_A= m_I$.

{\bf The Ion Cooling Simulation}
Ions are evolved in the experimental potential~\cite{Rav11}, from a random initial distribution, for elastic only and the elastic with RCx collisions. 
The ion$-$atom binary interaction potential~\cite{CRCHBk,Ols69} determines the specifics of the scattering event. All collisions are instantaneous and the Poisson distribution which determines the time between collisions is adjusted so that the experimental reality and computational constraints are balanced. Each collision occurs with an impact parameter $b$, where $0<b<b_{max}$ and $b_{max}$ is determined by a combination of $\mathbf c$ and $\rho_0$ such that  $\theta_c> 60\mu$ radians. When $b\leq b_{cx}$, where $b_{cx}$ is the critical impact parameter for charge exchange, the average probability for RCx is $1/2$ and for $b>b_{cx}$ the probability of RCx rapidly falls to zero, where $b_{cx}$ is determined from the Rb$^+-$Rb molecular potential~\cite{Ols69,Hol52}. The collision results in the change of ion velocity from $\mathbf c \rightarrow \mathbf c'$. The kinematics of the collision are computed following the treatment in McDaniel~\cite{McD64}.

\section{Acknowledgements}

The authors gratefully acknowledge T. Ray for useful discussions and RRI workshop and RAL for technical support.

\end{document}